\documentclass[9pt,twocolumn,twoside]{osajnl}
\journal{ol} 
\usepackage{siunitx}
\setboolean{shortarticle}{true}
\usepackage{color}
\title{Topological photonic crystal fibers based on second-order corner modes}
\author[1]{Ruirong Gong}
\author[1,*]{Ming Zhang}
\author[1]{Haibin Li}
\author[2,*]{Zhihao Lan}
\affil[1]{College of Science, Zhejiang University of Technology, Hangzhou 310023, China}
\affil[2]{Department of Electronic and Electrical Engineering, University College London,
	Torrington Place, London WC1E 7JE, United Kingdom}
\affil[*]{cim2046@zjut.edu.cn; z.lan@ucl.ac.uk}

\begin{abstract}
Photonic crystal fibers represent one of the most active research fields in modern fiber optics. The recent advancements of topological photonics have inspired new fiber concepts and designs. Here, we demonstrate a new type of topological photonic crystal fibers based on second order photonic corner modes from the Su-Schrieffer-Heeger model. Different from previous works where the in-plane properties at $k_z=0$ have been mainly studied, we find that in the fiber configuration of $k_z>0$, a topological bandgap only exists when the propagation constant $k_z$ along the fiber axis is larger than a certain threshold and the emergent topological bandgap at large $k_z$ hosts two sets of corner fiber modes. We further investigate the propagation diagrams, propose a convenient way to tune the frequencies of the corner fiber modes within the topological bandgap and envisage multi-frequency and multi-channel transmission capabilities of this new type of fibers. Our work will not only have practical importance, but could also open a new area for fiber exploration where many existing higher-order topological photonic modes could bring exciting new opportunities for fiber designs and applications.
\end{abstract}

\begin{document}
\maketitle
\par Photonic crystal fibers (PCFs) \cite{Russell03Sci, Knight03Nature}, which guide light by using a wavelength-scale periodic microstructure running along the entire length of the fiber, have the outstanding ability to overcome the limitations of conventional fibers and thus represent one of the most active research fields in modern fiber optics. The existence of a large variety of two-dimensional (2D) photonic crystals (PCs) in the cross section and the feasibility to integrate novel and functional materials with PCFs, have enabled light propagation via diverse guidance mechanisms and revolutionized the optical fiber technology. PCFs have proven to be useful in many important scientific and technological applications spanning a broad range of disciplines, such as nonlinear optics, material and laser science, sensing and spectroscopy \cite{Jr10RPP, Markos17RMP}.
\par Recently, the study of topological photonics \cite{Ozawa19RMP} exploiting topologically protected photonic modes to manipulate light in a way robust and immune to scattering loss and structural disorder has provided unique opportunities for new fiber concepts and designs. For example, in conventional fibers, bending loss and propagation attenuation due to scattering by unavoidable imperfections of the structure, have limited the performance of fiber devices \cite{Markos17RMP}. The recently proposed one-way fiber modes protected by the second Chern number in magnetic Weyl PCs \cite{Lu18NCfiber} and one-way surface modes robust against backscattering by defect and surface roughness in magneto-optical coaxial waveguides \cite{Wang20JJAP} have offered new avenues to overcome these limitations and greatly improve the fiber performance. Moreover, novel fiber concepts inspired by topological photonics, such as, Dirac mode induced light trapping and guidance in the absence of photonic bandgap and total internal reflection \cite{Xie15Light}, bianisotropic waveguides with nontrivial topological structure in momentum space \cite{Klimov18PRBbianisotropic}, topological Bragg fibers induced by Aubry-Andre-Harper cladding modulation \cite{Pilozzi20OL}, hybrid topological photonic localisation of light \cite{Makwana20OE} and Kekule modulation enabled Dirac-vortex fibers \cite{Lin20Light}, have been proposed and offered unprecedented opportunities in fiber designs and applications.
\par The aim of this Letter is to introduce a new type of topological photonic crystal fibers (TPCFs) based on second-order topological photonic modes. Higher-order topological phase \cite{Schindler18SciAdvHOTI, Xie20Review} is a recently discovered new type of topological phase, which features  topologically nontrivial boundary modes that are more than one dimension lower than the bulk. In the emerging field of higher-order topological photonics \cite{Kim20LightReview}, the most widely explored case is topological corner modes in 2D photonic systems based on Su-Schrieffer-Heeger (SSH) models \cite{Xie18PRBcorner,Chen19PRLcorner, Xie19PRLcorner,Kim20Nanophotonics}, which essentially are highly localized 0D states and have found promising applications in topological nanolasers \cite{Han20ACSpho, Kim20NCcornerlasing, Zhang20Lightcornerlaser}, cavity quantum electrodynamics \cite{Xie20LPRcorner} and high-quality factor nanocavity \cite{Ota19OpticaCorner}. However, the localized photonic corner modes explored in the literature up to now are non-propagating because most systems are 2D and the focuses have been placed on in-plane properties. Though fibers can be taken as approximation of 2D structures because they are effectively infinite along the fiber axis, one cannot take them as 2D by setting the wave vector $k_z$ along the fiber axis to zero. In other words, while the corner states are well-confined in the transverse plane, it is crucial to study the fiber properties at $k_z>0$ in order to understand propagating effects of the corner states along the fiber axis, which is the main motivation of this study. We find interestingly that due to the mixing of transverse magnetic and electric modes when $k_z\neq 0$, increasing $k_z$ beyond a certain threshold can induce a topological bandgap for both the two bands below it, which implies that there would exist two sets of propagating corner fiber mode at a  properly designed corner. We further investigate the propagation diagrams, demonstrate a convenient way to tune the frequencies of the corner fiber modes within the bandgap and envisage multi-channel transmission capabilities of this new type of TPCFs. Our work will open a new area of research where higher-order topological photonic modes could bring exciting new opportunities for fiber designs and applications.

\begin{figure}[t!]
	\centering
	\includegraphics[width=0.8\linewidth]{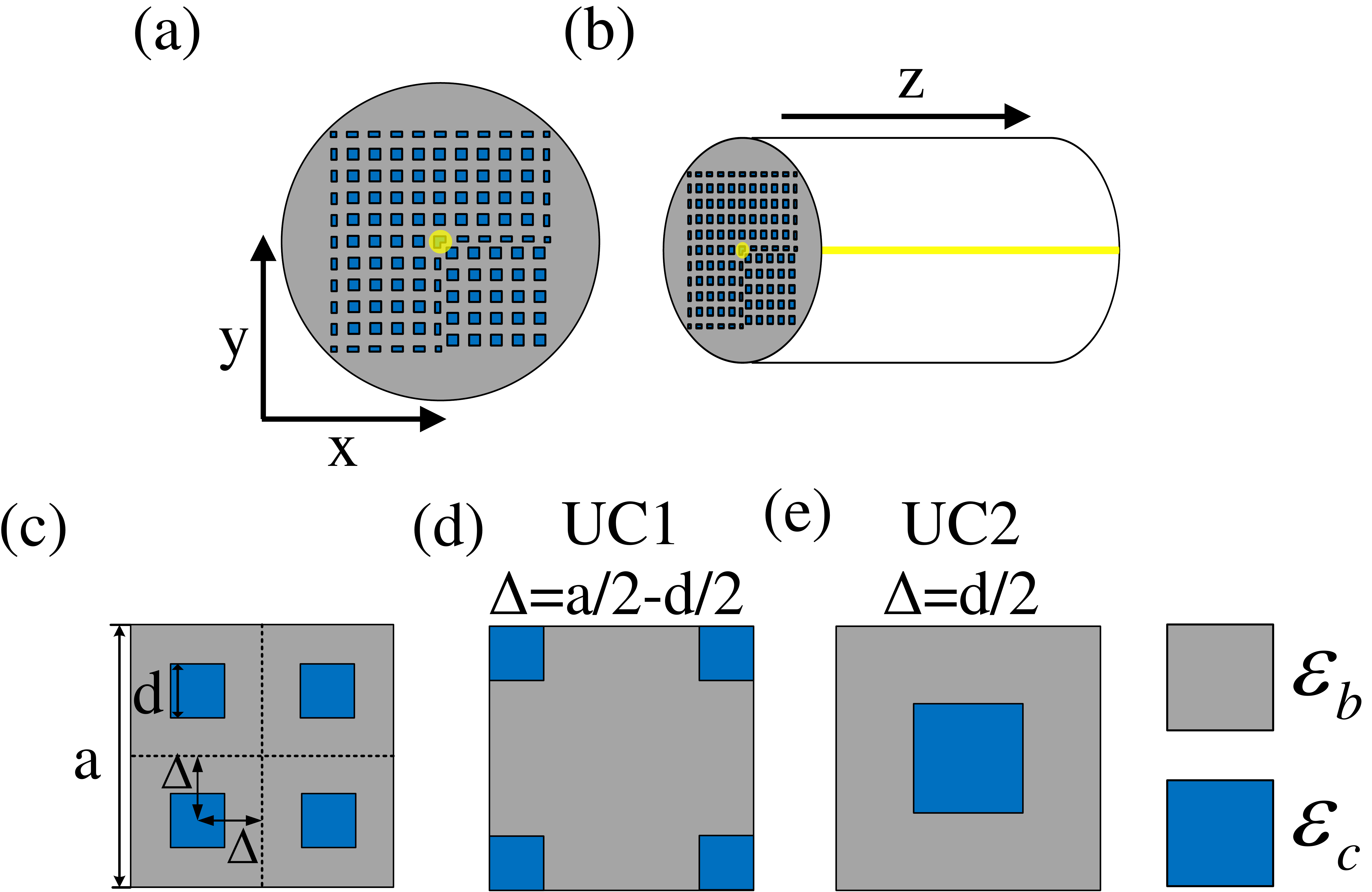}
	\caption{\label{fig:fig1}Schematics of the x-y cross sectional view (a) and the side view (b) of a topological photonic crystal fiber along z direction based on second order corner modes, where the yellow dot and line in (a) and (b) indicate the location of the corner fiber modes. (c) Unit cell of the photonic crystal in the x-y cross section with lattice constant $a$, which contains four square dielectrics with side length $d$ and dielectric constant $\epsilon_c=12.96$ immersed in a silica glass background with dielectric constant $\epsilon_b=2.1$. The four square dielectrics are located at ($\pm \Delta$, $\pm \Delta$) with $\Delta=a/2-d/2$ for (d) and $\Delta=d/2$ for (e). }
	\noindent\rule[0.25\baselineskip]{\linewidth}{0.5pt}
	\vspace{-9mm}	
\end{figure}

\par The proposed TPCFs based on second-order corner modes are shown schematically in 
Figs.\ref{fig:fig1}(a) and (b). The PC structure, which is based on the 2D SSH model \cite{Xie18PRBcorner, Chen19PRLcorner, Xie19PRLcorner, Kim20Nanophotonics}, contains two regions that intersect at a $90^\circ$ corner marked by the yellow dot or line in Fig.\ref{fig:fig1}(a) or (b), along which the corner fiber mode can propagate effectively. The unit cell (UC) of the PC is given in Fig.\ref{fig:fig1}(c), which contains four square dielectrics with side length $d$ immersed in a background of silica glass ($\epsilon_b=2.1$). To demonstrate the main idea of corner fiber modes, we use hydrogenated amorphous silicon (a-Si:H) with a dielectric constant of $\epsilon_c=12.96$ at 1550 nm as the squares inclusions (similar solid-core PCF has been experimentally demonstrated in \cite{Healy11OEsi} and for more state-of-the-art solid-core PCF fabrication technologies, see \cite{Markos17RMP}), such that a sizable bandgap could be more conveniently formed due to the large dielectric contrast between a-Si:H and the background of silica glass. 
\par A previous study based on the 2D SSH model \cite{Liu17PRLzeroberry} suggests that a topological phase transition could be induced by tuning the intracellular to intercellular coupling ratio, which could be realized for the UC in Fig.\ref{fig:fig1}(c) by just simply expanding or shrinking the four dielectric inclusions towards the corners or center of the UC \cite{Xie18PRBcorner}, as illustrated in Figs.\ref{fig:fig1}(d) and (e) with the resulting two UCs denoted as UC1 and UC2 respectively. Due to the different symmetry properties of the bands for the two UCs in Figs.\ref{fig:fig1}(d) and (e), it was found \cite{Xie18PRBcorner} that while the first bandgap of the UC in Fig.\ref{fig:fig1}(d) is topologically nontrivial, that of Fig.\ref{fig:fig1}(e) is trivial and as such, corner modes at a $90^\circ$ corner between two regions built from UC1 and UC2 could emerge. However, as these previous results \cite{Liu17PRLzeroberry, Xie18PRBcorner, Chen19PRLcorner, Xie19PRLcorner, Kim20Nanophotonics} were obtained for the in-plane properties of 2D PCs with $k_z=0$, it is not clear for a fiber configuration at $k_z>0$, whether a topological bandgap hosting nontrivial corner fiber modes could exist. 
\begin{figure}[t!]
	\centering
	\includegraphics[width=1\linewidth]{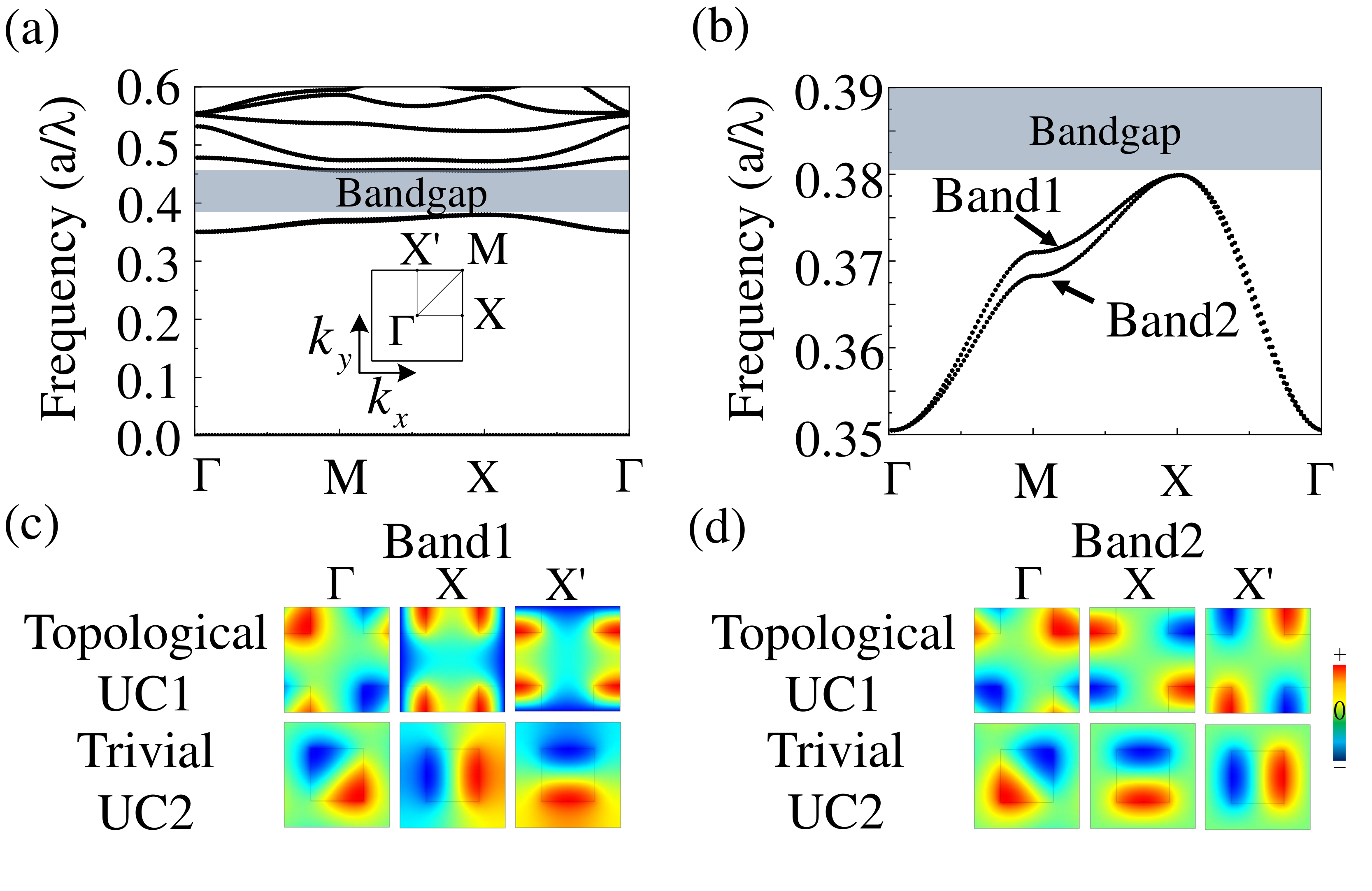}
		\caption{\label{fig:fig2}(a) The band diagram for the two unit cells, UC1 and UC2 in Figs.\ref{fig:fig1} (d) and (e) at $k_z=1.5\pi/a$ and $d=0.25a$, where the insert illustrates the first Brillouin zone with high symmetry points $\Gamma$, $X$, $M$ and $X'$ labeled. (b) shows the existence of two almost degenerate bands below the bandgap. (c) and (d) are field profiles ($Ez$) of the two bands below the bandgap at $\Gamma$, $X$, and $X'$ for UC1 and UC2. For both bands below the bandgap, the parities of the field profiles at ($\Gamma$, $X$, $X'$) are $(-1,+1,+1)$ for UC1 and $(-1,-1,-1)$ for UC2, indicating the topological nature of UC1 and trivial nature of UC2.}
	\noindent\rule[0.25\baselineskip]{\linewidth}{0.5pt}	
\end{figure}
 
\par  To demonstrate the feasibility of the corner modes to act as fiber modes, we show in Fig.\ref{fig:fig2} (a) the band structure of the two UCs of UC1 and UC2 in Figs.\ref{fig:fig1} (d) and (e) at $k_z=1.5\pi/a$, from which one can see a bandgap is indeed induced by $kz$.  It is to be noted that as UC1 and UC2 are different choices of the same PC, they have the same band structure as shown in Fig.\ref{fig:fig2} (a). Nonetheless, the symmetry properties of UC1 and UC2, especially the parities of the bands at high symmetry points $\Gamma$, $X$ and $X'$ of the first Brillouin zone (FBZ) can be different for the two UCs, a signature of different topology for UC1 and UC2. An interesting feature of the band diagram in Fig.\ref{fig:fig2} (a) is that there are two bands below the bandgap, which is very different from the band diagrams at $k_z=0$ \cite{Xie18PRBcorner, Chen19PRLcorner, Xie19PRLcorner, Kim20Nanophotonics, Han20ACSpho,  Kim20NCcornerlasing, Zhang20Lightcornerlaser, Xie20LPRcorner, Ota19OpticaCorner}. When $k_z=0$, the eigenmodes of 2D PCs can be separated to transverse magnetic (TM) and transverse electric (TE) modes and for practical applications, the bandgap is only designed for either TM or TE modes and as such there is only one band below the first bandgap. A common bandgap for both the TM and TE modes does not exist in the 2D SSH model \cite{Chen21PRApp}. 
However, when $k_z>0$, the modes can no longer be separated to TM and TE modes and this mode mixing results in the appearance of two bands below the bandgap, see Fig.\ref{fig:fig2} (b), which is a unique feature of the fiber configuration.

\par For the corner fiber modes to exist, the bandgap in Fig.\ref{fig:fig2} (a) has to be topological for one choice of the two UCs in Figs.\ref{fig:fig1} (d) and (e). The topology of the bandgap could be characterized by the  
extended Zak phase ($\theta_x,\theta_y$) in 2D through the polarization $P_j=2\pi\theta_j$ (with $j=x,y$) defined by \cite{Liu17PRLzeroberry}  
\begin{equation}
P_j = \frac{1}{(2\pi)^2}\int_{\textrm{FBZ}} dk_xdk_y \textrm{Tr}[A_j(k_x,k_y)], 
\end{equation}
where $A_j(k_x,k_y)=i\langle \psi | \partial _{k_j} | \psi \rangle$ is the Berry connection with $\psi$ the periodic part of the Bloch function and the integration is over the FBZ. Due to the inversion symmetry of UC1 and UC2, the polarization $P_j$ could be analyzed by the  parties of the electric field ($E_z$) profiles at the high symmetry points of the FBZ. In general, if the parities at $\Gamma$ and $X$ have an opposite (or same) sign, then the bandgap is topological (or trivial) \cite{Liu17PRLzeroberry, Xie18PRBcorner}. From the field profiles presented in Figs.\ref{fig:fig2} (c) and (d), one can see that the parities of the field profiles at $\Gamma$ and $X$ for both bands below the bandgap have an opposite (or same) sign for UC1 (or UC2), indicating that the 2D polarization $(P_x, P_y)$ for both bands below the bandgap is $(1/2,1/2)$ for UC1 and $(0,0)$ for UC2. The coexistence of nonzero $P_x$ and $P_y$ implies the existence of a nontrivial corner charge defined by \cite{Xie18PRBcorner, Chen21PRApp}
\begin{equation}
Q_{\textrm{corner}} = 4P_xP_y.
\end{equation} 
As such, the topological corner charge for both the two bands below the bandgap is 1 for UC1 and 0 for UC2. Due to this special feature, one could expect the existence of two corner fiber modes within the bandgap at a $90^\circ$ corner between two PCs made by UC1 and UC2, respectively. 
\begin{figure}[t!]
	\centering
	\includegraphics[width=1\linewidth]{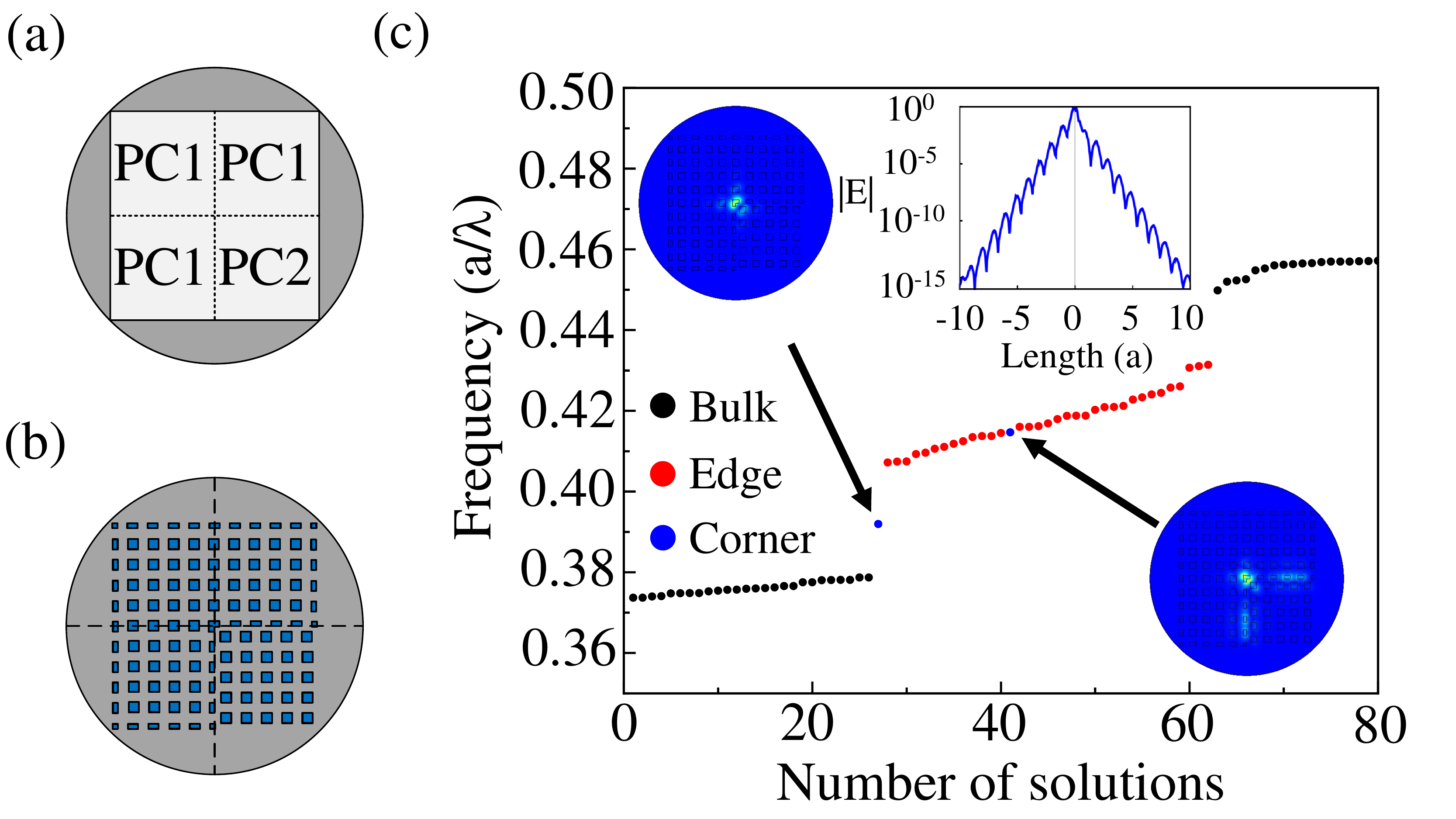}
	\caption{\label{fig:fig3}(a) Schematic cross section of a TPCF. (b) PC1 and PC2 are built from $5\times5$ UC1s and UC2s, respectively. (c) Eigenmodes of the fiber structure in (b) at $k_z=1.5\pi/a$ and $d=0.25a$, where apart from the bulk and edge modes, two corner fiber modes exist in the solutions whose field profiles are presented as inserts. Also shown in the insert is the logarithmic decay of the fiber mode field profile in a larger system.}
	\noindent\rule[0.25\baselineskip]{\linewidth}{0.5pt}
	\vspace{-9mm}	
\end{figure}
\par
\begin{figure}[t!]
	\centering
	\includegraphics[width=1\linewidth]{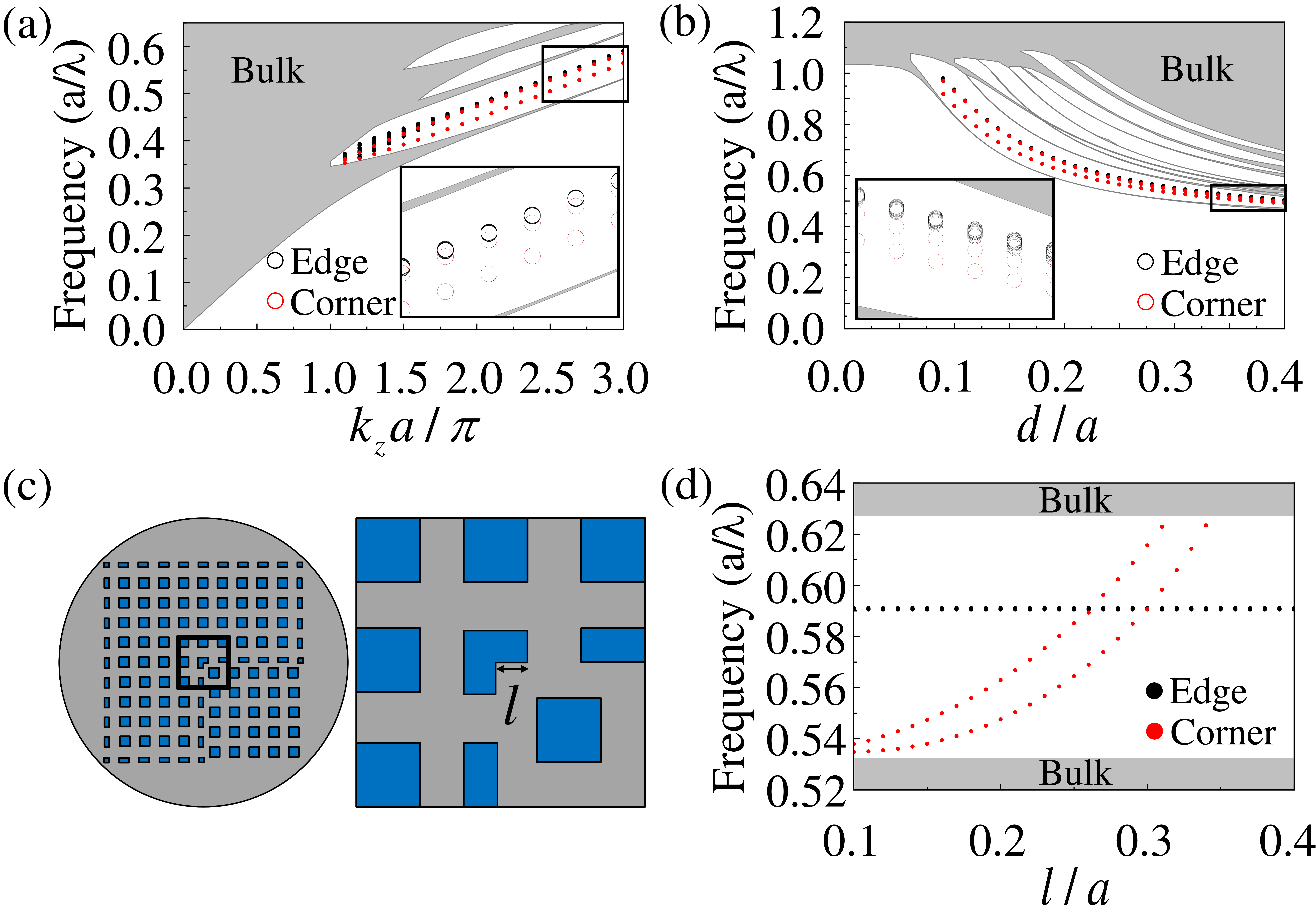}
	\caption{\label{fig:fig4} Band diagrams of the TPCF as a function of propagation constant $k_z$ at $d=0.25a$ (a) and  the side length $d$ of the square inclusions at $k_z=3\pi/a$ (b), where the inserts show the magnified parts marked by the  rectangles around the bandgap. (c) Varying the geometry of the block around the corner to tune the frequencies of the corner fiber modes, where a square of length $l$ is cut from the high dielectric inclusion. (d) The frequencies of the two fiber modes as a function of  $l$ indicated in (c) at $k_z=3\pi/a$ and $d=0.25a$.}
	\noindent\rule[0.25\baselineskip]{\linewidth}{0.5pt}
	\vspace{-5mm}	
\end{figure}

 To demonstrate this, we construct a square PC within the circular cross section of the fiber (see Fig.\ref{fig:fig3}(a)), which is occupied by three quarters of PC1 and one quarter of PC2 that are made of $5\times5$ UC1s and UC2s, respectively (see Fig.\ref{fig:fig3}(b)). The eigenmodes of the fiber configuration in Fig.\ref{fig:fig3}(b) at $k_z=1.5\pi/a$ are presented in Fig.\ref{fig:fig3}(c), from which one can see that apart from the bulk and edge fiber modes, there are two additional corner fiber modes with normalized frequencies of 0.392 and 0.415. While the energies of the bulk and edge fiber modes are distributed in the bulk and around the boundary between PC1 and PC2, the filed distributions of the corner fiber modes are localized around the corner between PC1 and PC2. It is to be noted that while the corner fiber mode at normalized frequency of 0.392 is highly localized around the corner (e.g., its field profile decays about 10 orders of magnitude from the peak value around the corner after a distance of 5 unit cells, see the insert of Fig.\ref{fig:fig3}(c)), the other corner fiber mode at 0.415 also has field distributions around the boundary between PC1 and PC2 due to the fact that this corner fiber mode is imbedded within the edge mode band. It is reasonable to wonder that the corner fiber mode at 0.415 could have a large loss and thus cannot serve as waveguiding fiber mode effectively. However, we will demonstrate later that the frequencies of the two corner fiber modes could be tuned continuly within the bandgap by varying the size of the block around the corner (see, e.g., Figs. \ref{fig:fig4}(c) and (d)), i.e., they cannot disappear by a small perturbation of the block size around the corner due to the topological nature of the corner fiber modes.

\par In order to have a better understanding of the TPCF performance, we map out the band diagrams of the TPCF when varying the propagation constant $k_z$ and the size $d$ of the square dielectric inclusions. First, the band diagram as a function of $k_z$ at $d=0.25a$ is presented in Fig.\ref{fig:fig4}(a), from which one can see that the bandgap for the corner fiber modes only exists when $k_z\gtrsim1$. Thus the physics discussed in this work is completely new compared to the previous works on second order corner modes carried out at $k_z=0$ \cite{Xie18PRBcorner,Chen19PRLcorner, Xie19PRLcorner,Kim20Nanophotonics, Han20ACSpho, Kim20NCcornerlasing, Zhang20Lightcornerlaser, Xie20LPRcorner, Ota19OpticaCorner}. As $k_z$ increases, one can also see from  Fig.\ref{fig:fig4}(a) that the bulk and edge mode bands become less dispersive as the couplings between the modes become weaker and consequently, the frequencies of the edge mode band within the bandgap gradually merge to a line. As such, the corner fiber mode imbedded in the edge modes at small $k_z$ (e.g., see Fig.\ref{fig:fig3}(c)) will be separated from the edge mode band at large  $k_z$, see the insert of Fig.\ref{fig:fig4}(a) for the zoomed in region at large $k_z$. Fig.\ref{fig:fig4}(b) shows the band diagram of the TPCF as a function of the size $d$ of the square inclusions at $k_z=3\pi/a$. As $d$ increases, the occupancy of the high dielectric material will increase and as such the frequencies of the edge and corner fiber modes decrease. As the insert of Fig.\ref{fig:fig4}(b) shows, in this case, the frequencies of the two corner fiber modes are separated from the edge mode band.

\par Considering the topological nature of the corner fiber modes and the fact that the field profiles of the corner fiber modes are localized around the corner, we propose an interesting way to tune the frequencies of the two corner fiber modes by varying the geometry of the dielectric block around the corner as illustrated in Fig.\ref{fig:fig4}(c), i.e., we cut off a square with length of $l$ from the center square dielectric inclusion, where $l=d/2$ corresponds to the cases in Figs.\ref{fig:fig4}(a) and (b) without modification. It could be expected that varying $l$ will only change the frequencies of the two corner fiber modes with negligible effect on the bulk and edge modes. The evolution of the eigenfrequencies of the modified fiber structure as a function of $l$ around the bandgap is shown in Fig.\ref{fig:fig4}(d), from which one can see that while the frequencies of the bulk and edge modes indeed keep invariant, the frequencies of the two corner fiber modes could be tuned from the bottom to the top of the whole bandgap. Especially, the frequencies of the two corner fiber modes could be easily separated from the edge modes by changing $l$, which could be very useful in practical applications. 
\begin{figure}[t!]
	\centering
	\includegraphics[width=1\linewidth]{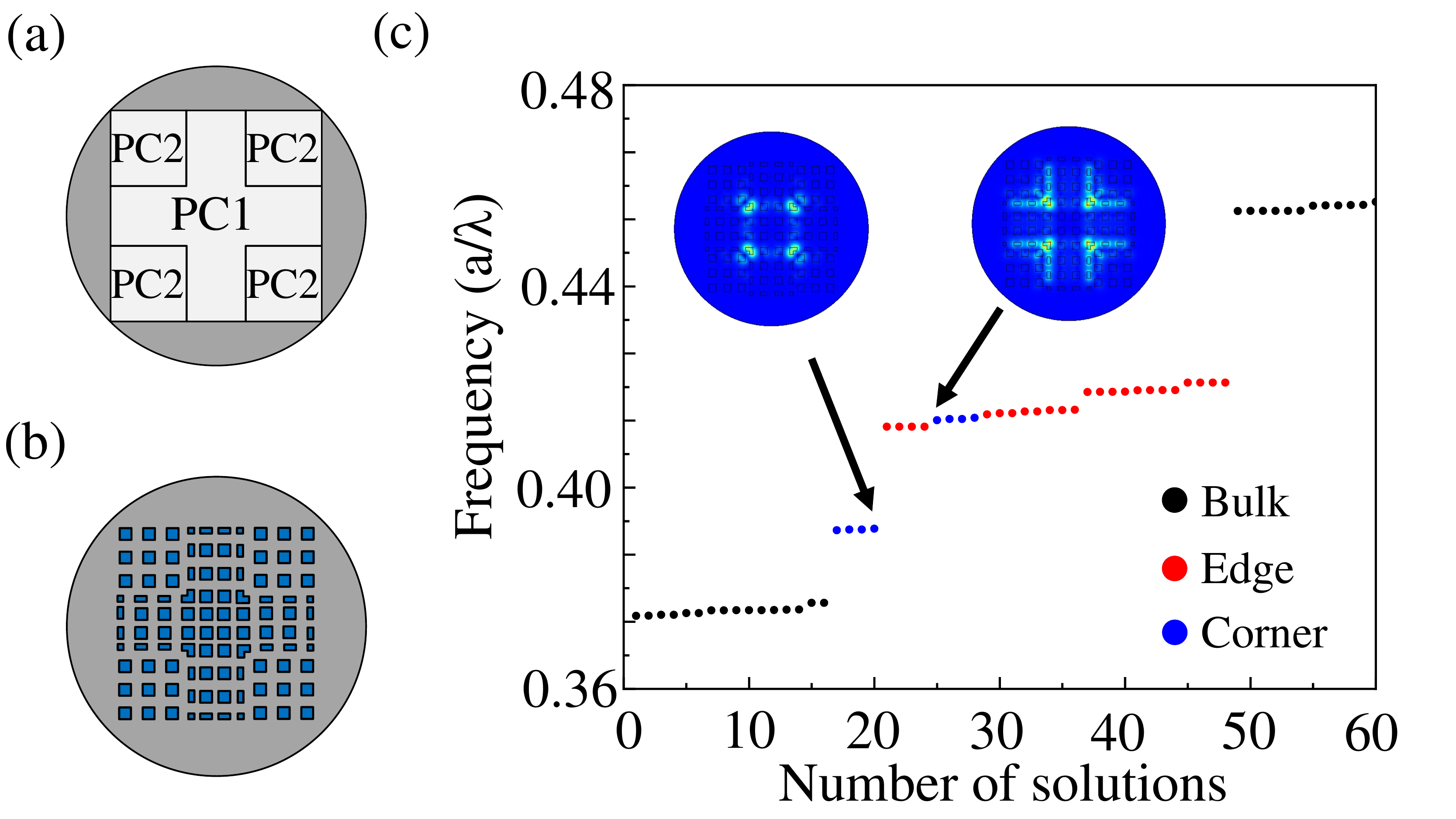}
	\caption{\label{fig:fig5}(a) Schematic for a scheme to achieve a multi-channel waveguide system based on multiple corners between PC1 and PC2. (b) A concrete realization where PC1 and PC2 in (a) are built from five $3\times3$ UC1s and four $3\times3$ UC2s. (c) Eigenmodes of the fiber structure in (b) at $k_z=1.5\pi/a$, which show the existence of two sets of corner fiber modes with each set containing four corner fiber modes. Typical field profiles of the two sets of corner fiber modes are shown as inserts. 
	 }
	\noindent\rule[0.25\baselineskip]{\linewidth}{0.5pt}
	\vspace{-9mm}	
\end{figure}

\par Finally, due to several prominent features of the proposed TPCFs, many interesting applications could readily be envisaged. Here, we demonstrate the design of a multi-channel waveguiding scheme based on multiple corners between PC1 and PC2 as illustrated in Fig.\ref{fig:fig5}(a), where a realization is presented in Fig.\ref{fig:fig5}(b) and the eigenmodes of this fiber structure at $k_z=1.5\pi/a$ are given in Fig.\ref{fig:fig5}(c). As expected, there appear two sets of corner fiber modes with each set containing four corner fiber modes due to the existence of four corners in the fiber design. Corner fiber modes with more complicated corner structures could  readily be envisaged and implemented. Furthermore, one can also tune the frequencies of the corner fiber modes in each corner separately using the method proposed in Figs.\ref{fig:fig4} (c) and (d) to realize multi-frequency and multi-channel light guiding. In practices, efficient excitation of the corner fiber modes could be achieved via either dipole sources~\cite{Han20ACSpho} or continuous lasers~\cite{Xie20LPRcorner} located around the corner with the same frequencies as those of the corner fiber modes. 

In conclusion, we have proposed a new type of TPCFs based on second order photonic corner modes from the SSH model. Different from previous works that have mainly focused on in-plane properties of the 2D SSH physics, we have revealed interesting fiber physics beyond the 2D configuration and found that a topological bandgap hosting propagating corner fiber modes only exists when the propagation constant $k_z$ along the fiber axis is larger than a certain threshold. We have further investigated the propagation diagrams, proposed a convenient way to tune the frequencies of the corner fiber modes within the bandgap and envisaged multi-frequency and multi-channel transmission capabilities of this new type of fibers. Our work could open a new area for novel fiber designs and applications via exploiting higher-order topological photonic modes. For examples, it would be interesting to explore the feasibilities of other PC structures that host corner modes for TPCFs, such as hexagonal \cite{Xie20NCcornerQSH} and kagome \cite{Li20NPkagomeCorner} lattices. The possibilities of TPCFs based on other types of higher-order topological photonic modes, such as quadrupole corner modes \cite{He20NCquadrupole} and hinge modes \cite{Benalcazar20arxivPhotHinge}, are also worthy of investigation. 

\medskip
\noindent\textbf{Disclosures.} The authors declare no conflicts of interest.

\end{document}